# Nanoscale electric-field imaging based on a quantum sensor and its charge-state control under ambient condition

Ke Bian[1,2,6], Wentian Zheng[1,6], Xianzhe Zeng[1], Xiakun Chen[1], Rainer Stöhr[2], Andrej Denisenko[2], Sen Yang[3], Jörg Wrachtrup[2*], Ying Jiang[1,4,5*]

[1]*International Center for Quantum Materials, School of Physics, Peking University, Beijing 100871, P. R. China*

[2]*3rd Institute of Physics, University of Stuttgart and Institute for Quantum Science and Technology (IQST), Pfaffenwaldring 57, D-70569, Stuttgart, Germany, Max Planck Institute for Solid State Research, Stuttgart*

[3]*Department of Physics, The Chinese University of Hong Kong, Shatin, New Territories, Hong Kong, China*

[4]*Collaborative Innovation Center of Quantum Matter, Beijing 100871, P. R. China*

[5]*CAS Center for Excellence in Topological Quantum Computation, University of Chinese Academy of Sciences, Beijing 100190, P. R. China*

**Nitrogen−vacancy (NV) centers in diamond can be used as quantum sensors to image the magnetic field with nanoscale resolution. However, nanoscale electric-field mapping has not been achieved so far because of the relatively weak coupling strength between NV and electric field. Using individual shallow NVs, here we succeeded to quantitatively image the contours of electric field from a sharp tip of a qPlus-based atomic force microscope (AFM), and achieved a spatial resolution**

---
[6] These authors contributed equally to this work: Ke Bian, Wentian Zheng.
* E-mail: j.wrachtrup@pi3.uni-stuttgart.de (J.W.); yjiang@pku.edu.cn (Y.J.)

**of ~10 nm. Through such local electric fields, we demonstrated electric control of NV's charge state with sub-5 nm precision. This work represents the first step towards nanoscale scanning electrometry based on a single quantum sensor and may open up new possibility of quantitatively mapping local charge, electric polarization, and dielectric response in a broad spectrum of functional materials at nanoscale.**

**Introduction**

The nitrogen-vacancy (NV) center, an atomic defect in diamond, is one of the most promising candidates for quantum computing[1,2], quantum information[3] and quantum sensing[4,5] under ambient condition, relying on the coherent manipulation of the spin ($S$=1) in its negatively charged state (NV⁻). Previous efforts have proven the shallow NVs to be powerful quantum sensors for detecting external signals, such as magnetic-[6-9], electric fields[10-12], strain[8] and temperature[13,14]. In particular, nanoscale magnetic-field imaging is possible by integrating a nano-diamond tip into the atomic force microscopy (AFM)[15], benefiting from the atomic size of NV. Recently, NV-based scanning magnetometry[16,17] based on the continuous wave-optical detected magnetic resonance (cw-ODMR) has been successfully used for the quantitative real-space imaging of superconducting vortex[18,19], multiferroic non-colinear order[20], skyrmion structure[21], and 2D magnetic materials[22].

In contrast, the electric-field sensing by the NV based on the Stark effect is much more difficult, and NV-based scanning electrometry has not been realized so far. The main reason arises from the relatively weak coupling strength of the NV with electric field, leading to small energy shifts of spin levels (typically <0.1 MHz)[10,12]. In this case, more advanced sensing schemes rather than the simple cw-ODMR are required, which strongly rely on the coherent properties of NVs[23]. However, the oscillating cantilevers of most non-contact AFM can modulate the distance between NV and target charged elements. Such a modulation induces electric noise, which would decrease the coherence time of the NV and degrades its sensitivity during the electric-field imaging[17]. Thus, it is ideal to bring the source of the external electric field as close as possible to the sensor while maintaining small oscillation amplitudes, which is challenging for the conventional NV-compatible AFM[18-22].

In this work, we circumvented this problem by using a qPlus-based AFM[24], which allows positioning a conductive tip close to the diamond surface below 1 nm with very small oscillation amplitudes (typically 100~300 pm). Thanks to the high field gradient of the sharp tip, we realized nanoscale quantitative imaging of external electric-field contours within the sensing regime of this atomic-size electrometer using pulsed-ODMR[25]. Furthermore, with the assistance of such high field gradients and photon ionization by the excitation laser, the local electric field of the tip can be applied to achieve control on the local electrostatic environment and charge state of shallow NVs with nanoscale accuracy (down to 4.6 nm), which may substantially enhance the coherence of NV and thus improve its field sensitivity.

**Results**

**Quantitative electric-field imaging at nanoscale**

The experimental setup is schematically shown in Fig. 1a (for details, see Methods). Shallow NVs with the depth of 5~10 nm were used as quantum probes, whose spin states were initialized by green laser and readout through red fluorescence. As a consequence of ion implantation for NV creation with subsequent annealing and acid-boiling cycles, a large number of defects were introduced both near the NVs and on the diamond surface[26,27]. Such defects can act as donors or acceptors, which are also considered to be the major source for decoherence and charge instability of shallow NVs[28,29]. In order to guide the conductive AFM tip towards a target fluorescent spot with high precision and to maintain a small tip-surface distance for achieving strong electric field, we chose the qPlus sensor working at the frequency modulation (FM) mode[24]. Because of its high stiffness and special asymmetric configuration, the qPlus sensor can work at very small tip heights (<1 nm) with tiny oscillation amplitudes (<100 pm) even under ambient conditions[30], thus hardly inducing electric noise to affect the coherence of NV⁻ when the tip is biased.

We carefully aligned the magnetic field (~9.6 Gauss) perpendicular to the NV axis according to hyperfine resonance spectra[10], such that the eigenstates of this spin-1 system change into an equal mixture of $m_s=\pm 1$ spin projection, i.e. $|\pm\rangle$ states (Fig. 1b). The strength of electric field can be measured from the frequency shift of $|\pm\rangle$ states due to the Stark effect[31]. Based on the pulsed-ODMR scheme (Fig. 1c) (for details, see Methods and Supplementary Fig. 1)[25], such a frequency shift ($\Delta f$) is clearly visible under different tip biases (Fig. 1d).

Next, we fixed the microwave frequency to be slightly off-resonant and monitored the fluorescence intensity during scanning the biased tip, allowing a single $\Delta f$ to be probed (Fig. 2a). In the scanning field-gradient image with the tip biased at -16 V, a triangle-like shape of field contour appears (Fig. 2b), which is consistent with the triangular-pyramid tip apex cut by focused ion beam (FIB) (See inset in Fig. 2a). However, at higher bias voltage and lower $\Delta f$, the electric field imaging changed into a nearly spherical structure (Fig. 2c), reflecting the overall symmetric shape of the tip shaft. From the preset microwave frequency, we can quantitatively estimate the field strength of the projected transverse component $E_\perp$ to be 119±3.3 kV cm$^{-1}$ at the ring position in Fig. 2b (for details of such quantitative estimation, see Supplementary Text 1). The corresponding electric-field gradient can be obtained from the line width of the ring structure and the minimum detectable field strength of the NV (~17.6 kV cm$^{-1}$) (see the arrows in Fig. 2b). Fig. 2d shows the simulated field distribution based on the model of tip with triangular shape (Fig. 2a). By adjusting the geometric parameters of the tip, we obtained the simulated field-gradient imaging (Fig. 2e), which agrees well with the experimental data (Fig. 2b).

From the field gradient in Fig. 2b, we can estimate that a tip amplitude of several nm will lead to a spectral broadening comparable to the intrinsic width of the pulsed-ODMR spectra (Supplementary Text 2). Therefore, the small amplitude (<100 pm) of our qPlus sensor is important to reduce the electric noise. It's worth noting that the best field resolution achieved in this work (~17 nm) (Fig. 2b) is determined not only by the absolute field gradient of the tip, but also the field sensitivity and size of the quantum sensor. Such a spatial resolution allows us to locate the AFM tip upon the NVs with high precision (Fig. 2f). We expect a resolution improvement by at least one order of magnitude if a spin-echo scheme with a synchronized A.C. bias voltage is applied[10,12]. Finally, using the geometric parameters obtained from the simulation, we deduce that the largest field reachable in our system is ~14 MV cm$^{-1}$ (Fig. 2d).

**Highly efficient charge-state control of NV by AFM tip**

In contrast to magnetic field imaging, measuring of charges and electric fields by NV centers can have a massive effect on the electronic structure of the NV center itself. In particular, the strong electric fields from the sharp tip can result in locally tuning the charge state of single NVs (see more details in Supplementary Fig. 2 and Supplementary Text 3). As shown in Fig. 3a, a positively biased tip (+90 V) was first positioned above a bright NV, and its fluorescence was slightly enhanced. However, when the tip bias was decreased to zero, this NV became non-luminescent. Its fluorescence revived again after applying a negative tip bias (-90 V) and remained stable afterwards at zero bias. Fig. 3b shows the fast switching between the bright and dark states of the NV, in prompt response to the positive bias voltages applied (+10 and +40 V). When ramping the bias voltage on the dark NV, a sharp transition threshold in the fluorescence counts was observed around +25 V (Fig. 3c). Such a transition is highly repeatable during multiple cycles of the bias ramp.

The change of the fluorescence state most likely arises from charge-state transitions of NVs[29]. To confirm this conclusion, we measured the photoluminescence (PL) of single NVs under 488-nm excitation laser to avoid the spectral overlap between background Raman peak and PL features (Supplementary Fig. 3). Under three selected positive biases, both the characteristic zero phonon line (ZPL) and sideband unambiguously point to distinct charge states (Fig. 3d). The dark state at the zero bias is attributed to NV$^+$ state[32,33], while the bright states at +35 V and + 100 V are attributed to NV$^0$ and NV$^-$ states [34], respectively. We note that the transition between NV$^0$ and NV$^-$ is more obvious at smaller laser powers (see Supplementary Fig. 4 and Supplementary Text 4). Strikingly, the NV$^+$ state remains stable in the absence of the external tip field, even under long-term laser illumination (Supplementary Fig. 5), which is in clear contrast to the previous work[32,33].

**The mechanism of charge-state control**

In order to gain deeper insights into the charge transition process, we systematically ramped the tip bias within a larger range (Fig. 4a). From the start, a high positive voltage (+150 V) was applied on a bright NV (NV$^-$) and ramped towards the negative voltage (blue curve in Fig. 4a). The fluorescence suddenly drops around +25 V (similar to Fig. 2c) and recovers around -50 V, which correspond to the NV$^0$→ NV$^+$ and NV$^+$→NV$^-$ transitions (see Supplementary Fig. 3, b and c), respectively. Along the opposite ramp (red curve in Fig. 4a), however, the photon counts show negligible variations. We excluded the electron injection and the ionization effect of air as the origin of NV charging/discharging[35-37], considering the fact that the NV$^+$→NV$^-$ transition occurs at both positive and negative biases, and that the charge state is still tunable even if the tip is lifted several tens of nanometers above the diamond surface.

We recall that the manipulated NV$^+$ does not relax back to NV$^-$ in the absence of the external electric field. In addition, the charge state of NV becomes tunable only after applying a sufficiently high positive bias voltage. Therefore, the NV$^+$ should be stabilized by the intrinsic surface polarization, which is induced by the strong tip field and preserves after the field is removed. Meanwhile, after turning off the laser, no charge transition was observed on the diamond surface even under large biases (±120 V). Therefore, we propose a model based on surface electron trapping arising from the concerted effect of photon-ionization and strong local electric fields.

Defects such as substitution nitrogen (P1 center) and vacancy complex act as deep donors and weakly dope the lattices surrounding the implanted NVs[27,38]. Recently, the existence of sp$^2$ carbon defects on the diamond surface with a density of > $4\times10^{13}$ cm$^{-2}$ caused by different surface treatments has also been confirmed by XNAES and DFT calculations[26]. As shown in Fig. 4b(i), the donors are easily ionized by the excitation of 488-nm or 532-nm laser due to its proximity to the conduction band minimum (~1.7 eV)[38]. Those ionized electrons are attracted towards the surface by the positively charged tip and trapped in the unsaturated carbon defects at the surface. Considering that those surface defect states appear 3.3~4 eV below the conduction band minimum[26] and the lack of mobile holes in the valence band, the trapped electrons are very stable under thermal equilibrium condition. Together with the ionized donors underneath, the charged surface establishes a large built-in electric field at zero tip bias (ii in Fig. 4b), depleting NV$^-$ to NV$^+$. Such a built-in electric field competes with the tip-induced downward band bending under the positive bias[39-41], hence enabling the charge-state transition as shown in Fig. 3c. When reversing the bias polarity, those surface-trapped electrons are repelled back into the bulk through photo-assisted tunneling process, and recombine with the ionized donors (iii in Fig. 4b). Consequently, the surface is depolarized along with the disappearance of the built-in electric field, leading to the revival of the NV⁻ state (iv in Fig. 4b). It's worth noting that we did not obtain NV⁺ up to a large negative bias (Fig. 4a), but only a small component of NV⁰ mixed into NV⁻ states (Supplementary Fig. 3c). This might arise from the less efficient screening of the negatively charged tip than that of the positive tip.

The built-in electric field induced by surface electron trapping can be further confirmed by local contact potential difference (LCPD) measurements[42] (see Supplementary Text 5 and Fig. 4c). We found that the contact potential of the tip in the polarized region is significantly more negative than that in the non-polarized region, consistent with that the polarized surface is negatively charged with trapped electrons. We also studied the role of surface chemistry in the charge-state manipulation. The charge states of NVs in two samples with different surface treatments (see Methods and Supplementary Fig. 6) are both tunable, except for a significantly larger positive threshold bias on the cleaner one (see Supplementary Fig. 7 and Supplementary Text 6). We conclude that the surface adsorbates can saturate the surface defects through charge transfer, leading to less efficient electron trapping and thus the decreased built-in electric field.

**Spatial precision of local charge-state control**

Finally, we demonstrate spatial control on the charge state of single NVs with unprecedented precision. In Fig. 5a, the randomly distributed bright NV⁻ can be selectively switched to dark NV⁺. In Fig. 5b, we chose a NV dimer with a separation of

~170 nm, which is beyond the resolution of our confocal imaging. The charge state of NVs in this dimer can be well controlled individually. In order to explore the ultimate limit of spatial control on the charge state, we obtained a charge-state transition image of a single NV by continuously recording fluorescence as scanning a positively biased tip (Fig. 5c). The resulting disk-like structure (upper panel in Fig. 5d) is very similar to the field contour image obtained by pulsed-ODMR (Fig. 2b). The sharp transition edge (middle panel in Fig. 5d), which arises from the $NV^+/NV^0$ transition, yields a spatial resolution of ~4.6 nm for charge state control (lower panel in Fig. 5d). Such a high spatial resolution is attributed to the high field gradient of the tip apex and the fast charge transfer near NV in response to the external field.

**Discussion**

In conclusion, we achieved, for the first time, the nanoscale imaging of electric fields using the NV as a local electrometer under ambient condition. Clearly, the sensitivity of the current sensing technique is not as good as the most sensitive scanning e-field imaging techniques, such as scanning single-electron-transistor (SET)[43,44] and scanning quantum dot (QD) microscopy[45,46]. However, most of those techniques only work under low temperature or ultrahigh vacuum conditions. Another asset of NV-based scanning electrometry is the ability to quantitatively measure the electric field, which is very challenging for electrostatic force microscopy (EFM)[47,48] or kelvin probe force microscopy (KPFM)[49]. Furthermore, the spatial resolution of the scanning NV electrometry is not very sensitive to the characteristics of tip apex. Such a feature surpasses many scanning e-field probes, whose spatial resolution closely relies on the tip apex, such as the size of micro-sensors (scanning SET, scanning QD), the details of the exact tip termination (KPFM, EFM), etc.

Feeding A.C. electric field to the AFM tip may open up further possibility of exploring frequency-dependent surface dielectric response at nanoscale based on ultrasensitive coherent measurements. By tuning the local electrostatic environment such as appropriately depleting the surrounding charges of NV in a controllable manner, our technique is promising to reduce the spin/charge noise and enhance the spin coherence/contrast of the shallow NVs[50], which can improve the electric-field sensitivity even up to single elementary charge. Furthermore, integrating the NV onto the scanning tip may become an emerging tool in nanotechnology and opens up new possibilities of probing the local charge and electric polarization in a broad spectrum of functional materials, such as solar cells, ion batteries, ferroelectrics, multiferroics, electronic devices, etc. However, the NV inside the diamond tip after complex fabrication process suffers from low coherence time, low spin contrast and poor charge stability, which limit the sensitivity of diamond tips and its application for electric-field measurement. Recently, the nanometer-sized diamond pillar array with a flat top has been successfully produced and the sensitivity of NV is comparable to ones inside non-structured diamonds. The next step would be transferring and integrating the single diamond pillar onto the AFM tip.

In addition, we have also demonstrated that the electric control on the charge-state transition of NV can be realized with sub-5 nm precision. The manipulated charge states are intrinsically stable under thermal equilibrium condition, which is, in principle, applicable for other kinds of solid-state qubits in diamond[51,52] or silicon carbide[53]. Those results pave the way to construct complex qubit network for scalable quantum register and quantum processor, especially on samples containing ensemble qubits such as NVs. The strong local field and the small capacitance from the sharp tip can greatly enhance the transfer speed of charge carriers and enable the fast charging/discharging of NV within microseconds, which is vital for nuclear-based quantum storage[33] and spin to charge readout[54,55].

**Methods**

**Experimental setup.** All the data in this work were recorded in our home-built scanning probe microscope (SPM) system, which was specially designed for achieving excellent compatibility with nitrogen-vacancy (NV) center technology. The SPM part includes a compact Pan-type scanner head[56], integrated with vector magnets and high-frequency transmission cables. An oil immersed objective (*N. A.*=1.3) was used for photon collection, and the focus spot was driven by a commercial piezoelectric scanner (Physik Instrumente). We chose two types of qPlus sensors (one with $k$=1.8 kN m$^{-1}$, $Q$~2000, $f$=33 kHz, the other with $k$=3.6 kN m$^{-1}$, $Q$~2500, $f$=53 kHz) equipped with a tungsten tip (25 μm in diameter) for atomic force microscopy (AFM) measurements[15]. The tungsten tip was first electrochemically etched in the NaOH solution, followed by cleaning and sharpening with FIB. The bias voltage was supplied by an AO output of NI-DAQ (National Instrument) and amplified through a commercial voltage amplifier (-150 V ~ +150 V, CoreMorrow). The bias was applied on the tip, and the ground reference was on the waveguide and shielding box of the SPM scanner. The details of the NV-combined SPM will be published elsewhere. Single NV centers were located by our home-built confocal imaging system. The spin state of NV⁻ was initialized by a 532-nm laser and readout by fluorescence lights. The 532-nm laser was chopped by an AOM (Acoustic-Optic modulator, Gooch & Housego) in the double-pass working mode and subsequently shaped by a single-mode fiber. A half-wave plate was used for adjusting the polarization of input laser. Fluorescence photons through 650-nm long-pass (LP) filter were ultimately collected by APD (Avalanche Photodiode, Excelitas). A 100-μm multi-mode fiber was connected to APD and used as the pinhole in our confocal setup. We used the NI-DAQ for counting photons. Microwaves (MW) were generated by a Keysight signal generator (N5181B), chopped by switchers, amplified by a Minicircuit power amplifier, and then fed through on-chip waveguide for flipping the electron spin of $NV^-$. The synchronization of laser pulse, microwave pulse and counter timing was executed through a multi-channel-digital pulse generator (PBESR-500, Spincore). The AFM data acquisition was realized by a commercial Nanonis package, while the photon counts and coherent manipulation were realized by the self-programmed Labview VIs. Photoluminescence (PL) spectra were recorded by a spectrometer from Princeton Instrument (SP2300). When performing PL measurements, the 532-nm laser was replaced by a 488-nm one and the 650-nm LP was replaced by a 550-nm one to avoid the spectral overlap between background Raman peak and PL features of $NV^0$. The typical acquisition time of PL spectra is 2~30 minutes per curve depending on the signal-to-noise ratio under excitation laser of 400~500 μW, during which the charge state of NV was not changed.

**Pulsed-ODMR and nanoscale electric-field imaging.** For pulsed-ODMR measurements, we applied a microwave π-pulse for flipping the electron spin (Supplementary Fig. 1a), during which no laser was illuminated on the NV. In this case, the detection sensitivity of external electric field is largely enhanced compared to continuous-wave ODMR (Supplementary Fig. 1b), because of the suppressed power-broadening induced by continuous excitation laser and microwave. The spectral resolution depends on the duration time of π-pulse (Supplementary Fig. 1c), which is ultimately limited by the NV's coherence time[25]. In our experiments, the maximum duration of π-pulse is ~5.4 μs, leading to a spectral resolution of ~300 kHz, which corresponds to the minimum detectable field strength of ~ 17.6 kV cm$^{-1}$ and sensitivity of ~35.2 kV cm$^{-1}$ Hz$^{-1/2}$ (Supplementary Fig. 1d). For scanning field-gradient imaging[57], we fixed the frequency of microwave pulse several hundreds of kHz away from the resonance peak under the zero field. The resulting frequency shift reflects the strength of electric field, under which the spin resonance decreased photon counts. The tip-surface distance was controlled through oscillation amplitudes (100~300 pm) and frequency shifts (+10 ~ +70 Hz) of the qPlus sensor in frequency-modulation (FM) mode. The integration time per pixel is 2~4 s, leading to a total acquisition time of 2~4 hours per image (48 pixels × 48 pixels). All the fluorescence data for electric field sensing were recorded under the 532-nm excitation laser with the 650-nm long-pass (LP) filter. All the pulsed-ODMR data were normalized by $f_{sig}/f_{ref}$, where $f_{sig}$ and $f_{ref}$ are the averaged fluorescence photons within ~300-ns signal and reference counting-window defined in Supplementary Fig. 1d. A temperature control system was built for suppressing the thermal drift[58]. The temperature setpoint was maintained by 4 PID (proportional-integrating-differential) units, and the whole NV-SPM system was settled inside a thermal insulation box consisting of foams and acoustic-proof panels.

**NV creation and sample treatment.** The diamonds are commercial electronic-grade single-crystal chips purchased from Element Six. The intrinsic nitrogen concentration is below 5 ppb. The chips were milled into membranes with a thickness of 20~30 μm by laser cutting in DDK Inc. The diamond membranes were then implanted with 5-keV $N^{15}$ ions. A subsequent high-temperature annealing led to the diffusion of carbon vacancies, which were ultimately combined with $N^{15}$ donors. For sample A, in order to avoid complex process in ultraviolet (UV) lithography or electron beam lithography (EBL), we chose the etched copper films as a shadow mask during the evaporation of Cr/Au waveguide (Supplementary Fig. 6a). Because of the ~1-mm distance between the mask and diamond substrate, a small amount of Cr/Au was leaked into the waveguide gaps, leading to high confocal background (100~120 kcts s$^{-1}$ under 300 μW) and quenching of some NVs. After putting the chip into piranha solution and boiling it for 3 cycles, the confocal background was decrease to 10~20 kcts s$^{-1}$ under 300 μW and more NVs revived (Supplementary Fig. 6b). The AFM image shows that the surface of sample A is covered by 5-nm thick layers, which may correspond to the adsorbed water under ambient condition due to the hydrophilic termination of the surface after the acid treatment. For sample B, after the NV creation the chip was immersed in isopropanol solution for several months (Supplementary Fig. 6c). To keep the sample surface relatively clean, we didn't evaporate any waveguide and perform acid-boiling procedure on sample B. From the AFM image, the surface of sample B is free of the adsorbed layers as found on sample A (Supplementary Fig. 6d). Both the A and B chips were stuck onto a 170-μm silica substrate using UV-cured glues and inserted into our SPM system. The typical saturated fluorescence of single NVs in our system is 70~90 kcts s$^{-1}$ with a background of 10~15 kcts s$^{-1}$. The spin contrast of NVs on sample A measured through Rabi oscillation is 15%~25%, and the typical $T_2$ under spin-echo sequence is 15~30 μs.

**Data availability**

The data that support the findings of this study are available from the corresponding author upon reasonable request.

**Acknowledgements**

The authors thank Xinyu Pan & Gangqin Liu for the generous help on NV instrumentations, Zhihai Cheng for valuable suggestions on AFM technologies, Nan

Zhao & Marcus Doherty for insightful discussions on the data interpretation, and Bowen Song for valuable assistance on graphics and data processing. This work was financially supported by the National Key R&D Program under Grant Nos 2016YFA0300901, 2017YFA0205003, the National Natural Science Foundation of China under Grant Nos 11888101, 11634001, 21725302, the Strategic Priority Research Program of Chinese Academy of Sciences under Grant No. XDB28000000 and Beijing Municipal Science & Technology Commission under Grant No. Z181100004218006. R.S., A.D. and J.W. were supported by the ERC grant SMeL and VW Foundation as well as the EU via the projects ASTERIQS and QIA. S.Y. was supported by Hong Kong RGC (GRF/ECS/24304617, GRF/14304618 and GRF/14304419).

**Author contributions**

Y.J. and J.W. designed and supervised the project. K.B. and Y.J. constructed the experimental setup. R.S. and A.D. grew diamond chips and fabricated shallow NVs. K.B. and W.Z. performed experiments and data acquisition. X.Z performed finite element analysis. K.B, W.Z., X.Z., X.C., S.Y., J.W, and Y.J. performed data analysis and interpretation. K.B. and Y.J. wrote the manuscript with the inputs from all other authors. All the authors commented on the final manuscript.

**Competing interests**

The authors declare no competing interests.

**Figure legends:**

**Fig. 1 | Sensing the external field from AFM tip using single shallow NV. a**, Schematic graph of the home-built AFM system combined with NV technology. A metal tip was guided close to a specific NV using a qPlus sensor, where the electrode for bias is denoted by 1 and the electrodes for AFM signals are denoted by 2 and 3. A piezo actuator (grey arrow) is used for exciting the oscillations of qPlus. Microwave (MW) pulses feed through the on-chip waveguide. The transverse magnetic field $B_\perp$ is dentoed by a white arrow, with the magnitude of 9.6 Gauss. **b**, Left: The spin levels of NV⁻ under transverse magnetic field with varying strength. Hyperfine interaction of associated $N^{15}$ isotope is denoted by $A_\parallel$. The states ( $|1\uparrow\rangle$ , $|-1\downarrow\rangle$ ) and ( $|-1\uparrow\rangle$ , $|+1\downarrow\rangle$ ) (denoted by z-projection of the electron spin-1 and its associated nuclear spin-1/2) are equally mixed into $|+\rangle$ and $|-\rangle$, respectively. Right: Energy levels of $|\pm\rangle$ states under the electric field with varying strength. Arrows indicate the spin resonant transition under the microwaves ($f_+$ and $f_-$). **c**, Continuous wave-ODMR (upper panel) and pulsed-ODMR (lower panel) under zero electric fields. The hyperfine splitting cannot be resolved in continuous wave-ODMR (the grey shadow), but is clearly visible in pulsed-ODMR. The vertical dashed line denotes 2.87 GHz. **d**, Pulsed-ODMR spectra showing the electron-spin-resonance shift of $f_+$ under different tip biases. Curves are offset for clarity. AFM setpoint: $\Delta f_{AFM}$ =+20 Hz.

**Fig. 2 | NV-based nanoscale scanning electrometry and quantitative estimation. a**, Schematic graph showing the simple triangular-tip model with a spherical tip apex. $E_\perp$ and $E_\parallel$ (green lines) are projected onto a NV with a depth of $d$ at a tip height of $h$.

During the simulation, we set $\varepsilon_d$=5.7 for diamond material. Inset: Scanning electron microscope (SEM) image of FIB-treated tungsten tip. Scale bar: 1 μm. **b**, Scanning field-gradient imaging obtained at -16 V with ODMR setpoint of 700 kHz. The dashed arrows reflect the direction and magnitude of field gradient. **c**, Scanning field-gradient imaging of the tip under different biases. AFM setpoint: $\Delta f_{AFM}$ =+10 Hz. Pulsed-ODMR setpoint: $\Delta f$=400 kHz. **d**, Simulated field distribution along the surface normal ($E_z$) from the triangular-tip model (**a**) with a side length of 220 nm, apex radius of 30 nm, $h$=1 nm, $d$=5 nm and bias=-150 V. A field strength as large as 14.1 MV cm$^{-1}$ can be experienced by the NV. **e**, Simulated scanning-field gradient image at the bias=-10 V. The ODMR setpoint is set to be 700 kHz. The non-spherical shape arises from both the triangular tip and the fact that NV axis is not perpendicular to diamond surface. Scale bars in (**b**), (**d**) and (**e**) are 100 nm. **f**, A series of simulated scanning gradient-field images under the same geometric parameters as in (**e**), demonstrating a precision of 13 nm for positioning the NV underneath.

**Fig. 3 | Tip-induced charge-state transition of a single NV. a**, The process of charging the NV by local electric field from the AFM tip. Scale bar: 1 μm. (i) Confocal image including the target NV denoted by a dashed white ellipse before the charge state control. (ii) The tip with a high positive bias was guided close to the target NV, and the fluorescence of NV preserved. (iii) When the tip was grounded, the NV became completely dark. The small background denoted by red arrow arises from the tip-scattered light. (iv) Applying a high negative voltage to the tip switched the NV to bright state as before. (v) The bright state of NV remains when the tip was grounded

and retracted. **b**, Time-elapsed fluorescence shows the fast switch of NV charge state when the bias was set at +40 V or +10 V alternatively. **c**, Fluorescence-bias curves in the positive voltage range. Reproducible curves along with two cycles of bias ramps indicate the charge state control with high fidelity. **d**, PL spectra of the same NV in (**c**) under different biases, confirming the different charge states of NV. The blue and red arrows indicate the ZPL of $NV^0$ and $NV^-$, respectively. All the fluorescence data were recorded under the 488-nm excitation laser with the 550-nm LP filter.

**Fig. 4 | Schematic model of the charge state control. a**, Fluorescence-bias curves over the full-bias range. The dashed arrows denote the direction of bias ramp. **b**, Cartoon graphs (upper panels) and energy band diagrams (lower panels) showing the concerted effect of photon ionization and tip electric field. Upper panels: red and black balls indicate $NV^-$ and $NV^+$ states, respectively. Dashed arrows denote the local electric field. Lower panels: the purple curve denotes the donor's charge transition level, while blue and red short lines denote $NV^+/NV^0$ and $NV^0/NV^-$ transition levels, respectively. The waved green arrow denotes exciting photons. $E_F$, $E_C$ and $E_V$ denote the Fermi level, conduction and valance band edges of diamond, respectively. (i) Under a large positive bias, lots of photon-ionized electrons (dark blue balls) from the donors (cyan balls) are driven upward and trapped by the surface defects. (ii) A strong built-in electric field (light blue arrow) remains when the tip is grounded, leading to the upward band bending and the $NV^+$ state. (iii) The surface-trapped electrons are photoexcited and tunnel to the conduction band under negative bias, thence recombined with the donors, depolarizing the surface and leading to the $NV^-$ state again. (iv) When the bias is grounded again,

only shallow donors are depleted due to the electron transfer to surface defects and a small built-in electric field is formed, thus the NV⁻ state remains. **c**, LCPD measurements confirmed the negatively charged surface after polarization. Setpoint: $\Delta f_{\mathrm{AFM}}$ =+20 Hz.

**Fig. 5 | Local charge-state control with high spatial resolution. a**, Confocal image showing highly efficient charge-state control of arbitrarily selected NVs. The target NVs are denoted by black dashed circles. Scale bar: 1 μm. **b**, Charge state control of a NV dimer with a separation beyond the diffraction limit. The manipulated NV is denoted by 2. Scale bar: 0.6 μm. Upper panel: Line profile across the dimer and the corresponding Gaussian fit, yielding a separation distance of ~174 nm. Inset: Simulated confocal image of the NV dimer. Scale bar: 250 nm. **c**, Schematic graph demonstrating the basic principle of charge-state transition imaging. The AFM tip with a high field gradient is able to control the charge state of NV with high spatial resolution. The green shade denotes excitation laser, while the semi-circled curves denote electrostatic potential contour from the conducting tip. The dashed lines indicate the built-in electric field. **d**, Charge-state transition image of a single NV. Upper panel: A disk-like structure was obtained when scanning the metal tip with a high positive bias (+90 V). The edge of the disk reflects $NV^+/NV^0$ transition. Middle panel: The zoomed-in image as denoted by the black dashed rectangle in the upper panel. Lower panel: The profile along the red dashed line depicted in the middle panel. After subtracting a linear slope, the line profile was fitted by a step function, yielding a spatial resolution of 4.6 nm for charge-state control.

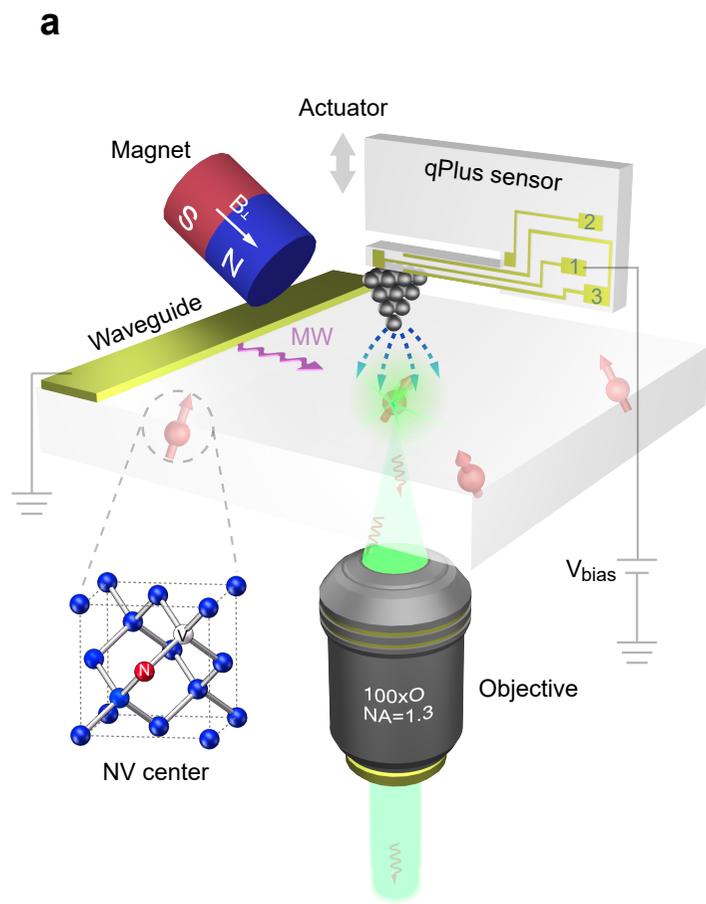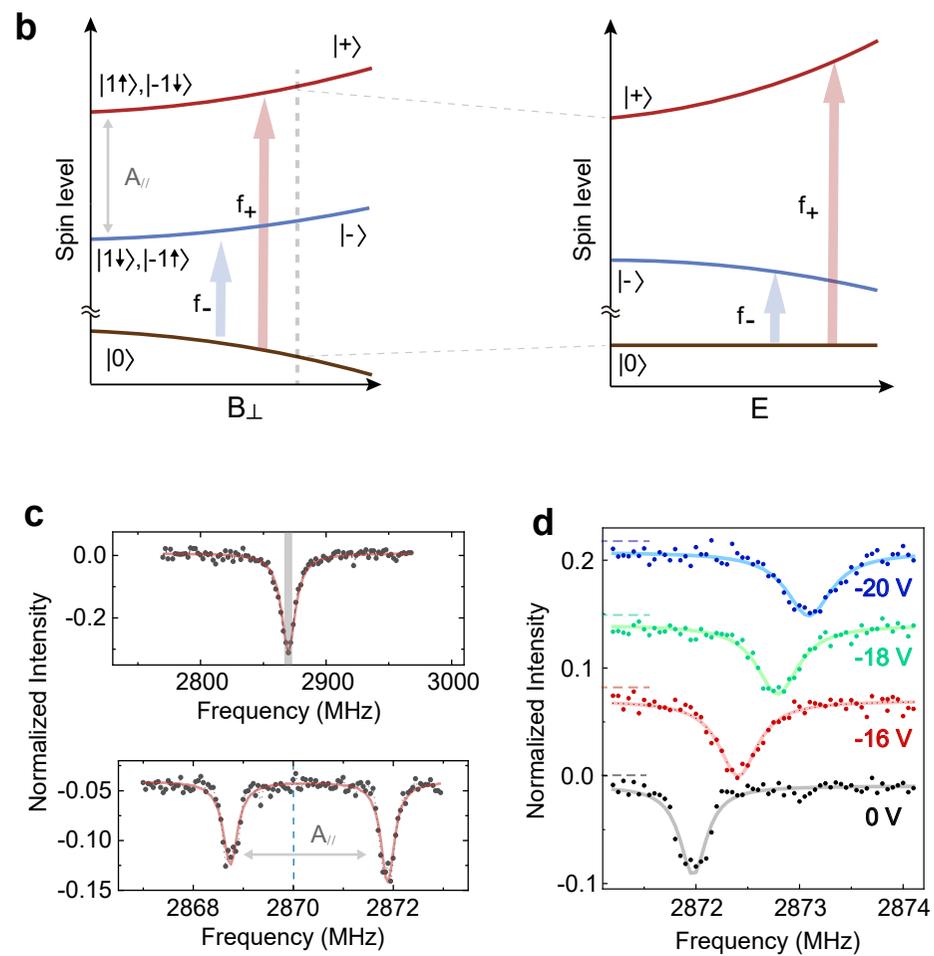

**Figure 1**

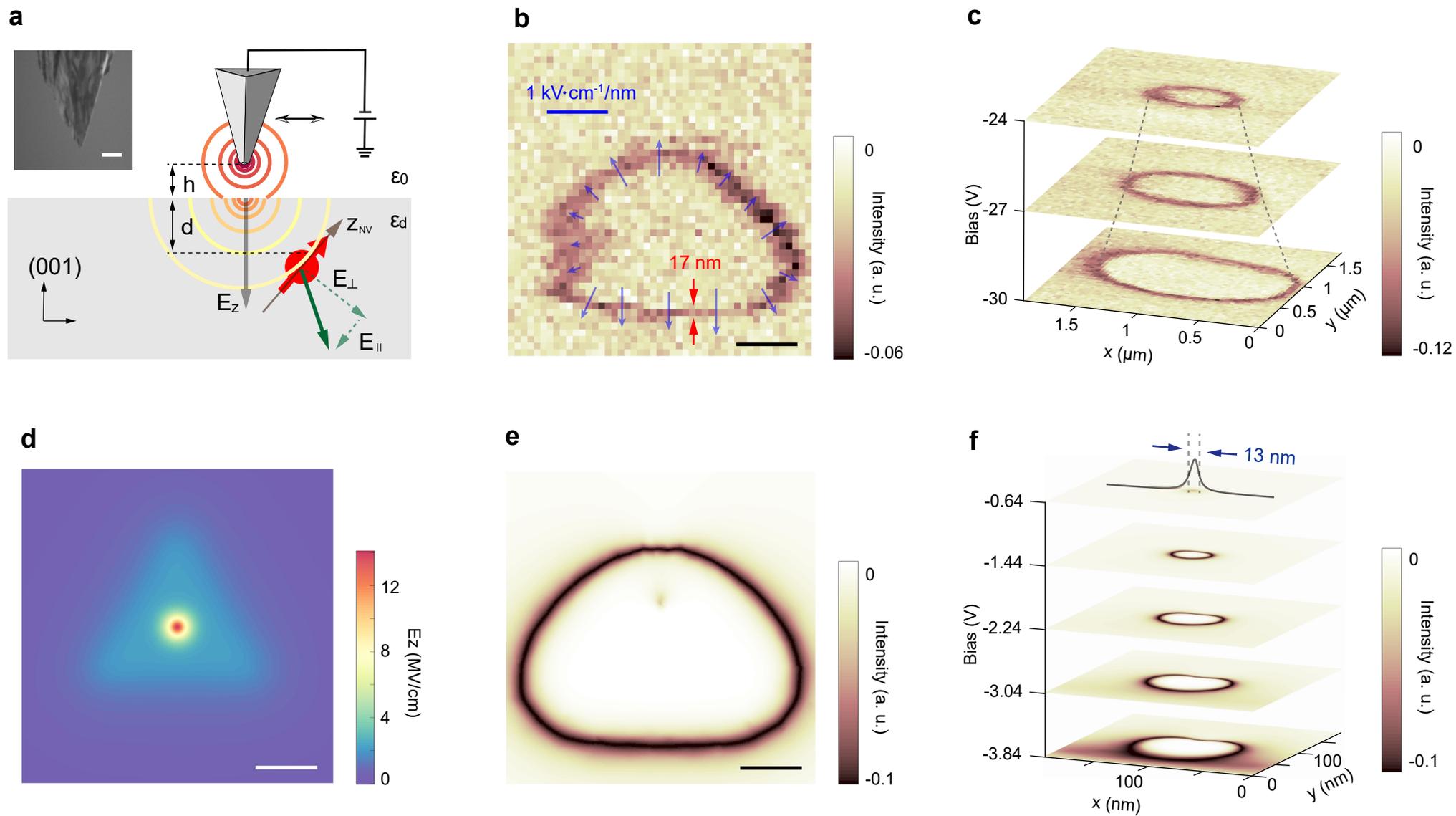

**Figure 2**

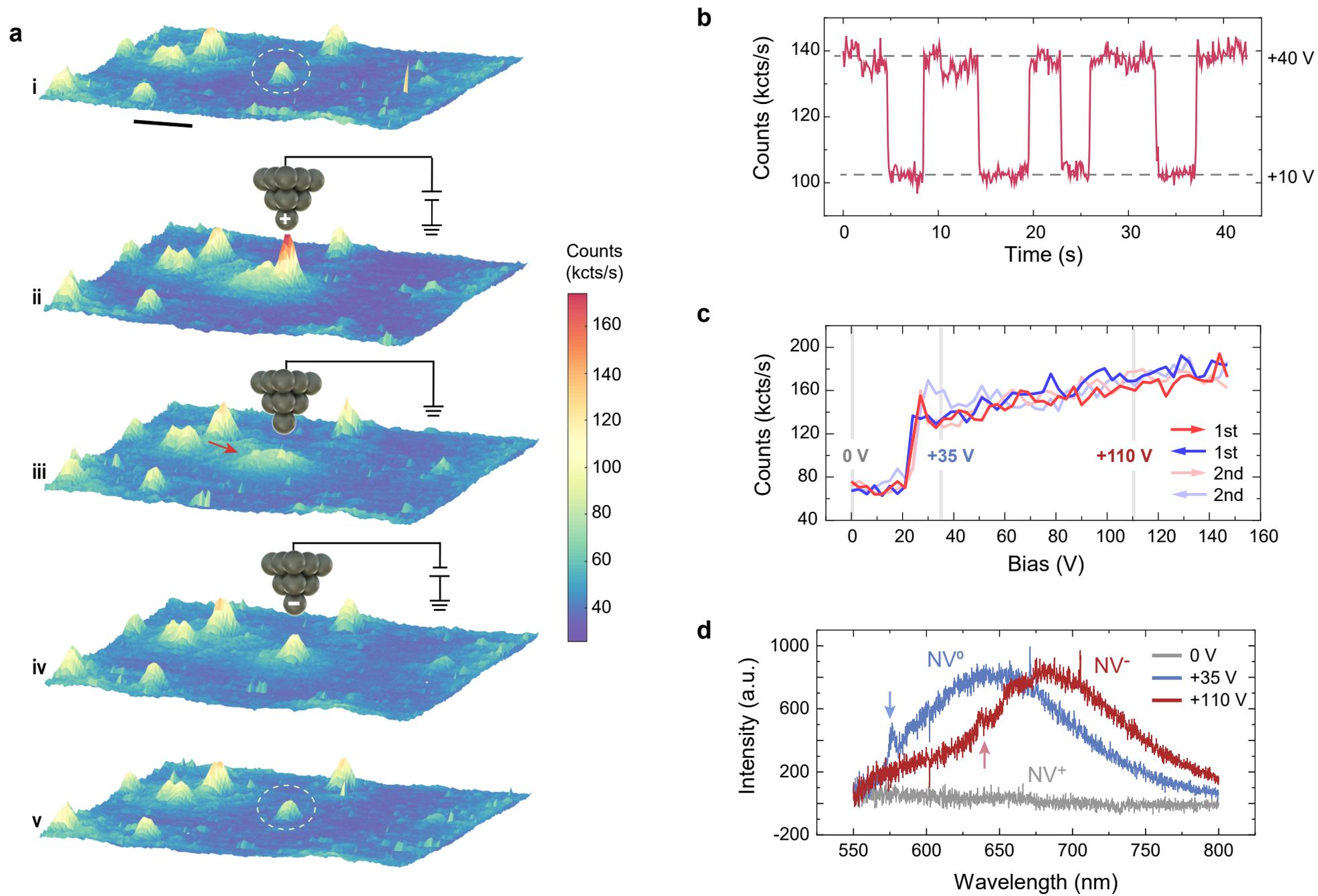

**Figure 3**

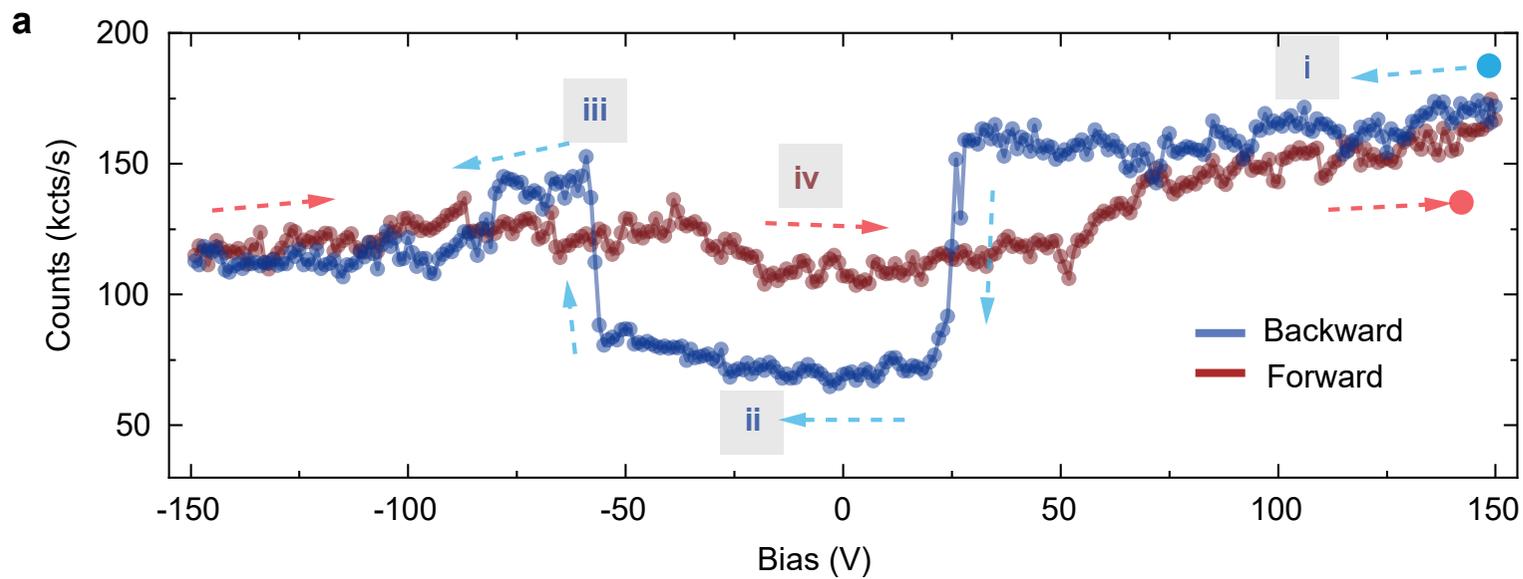
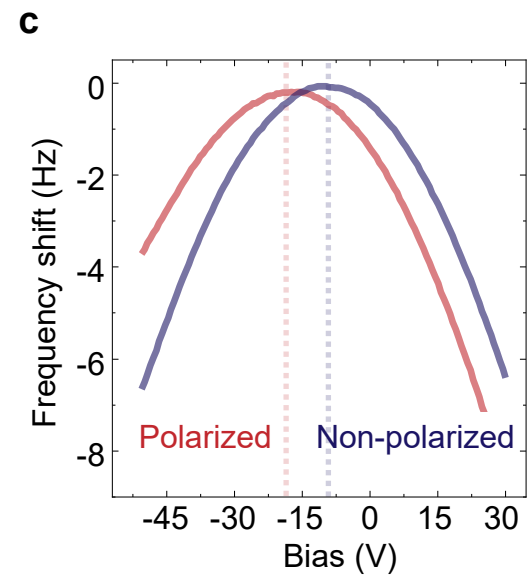
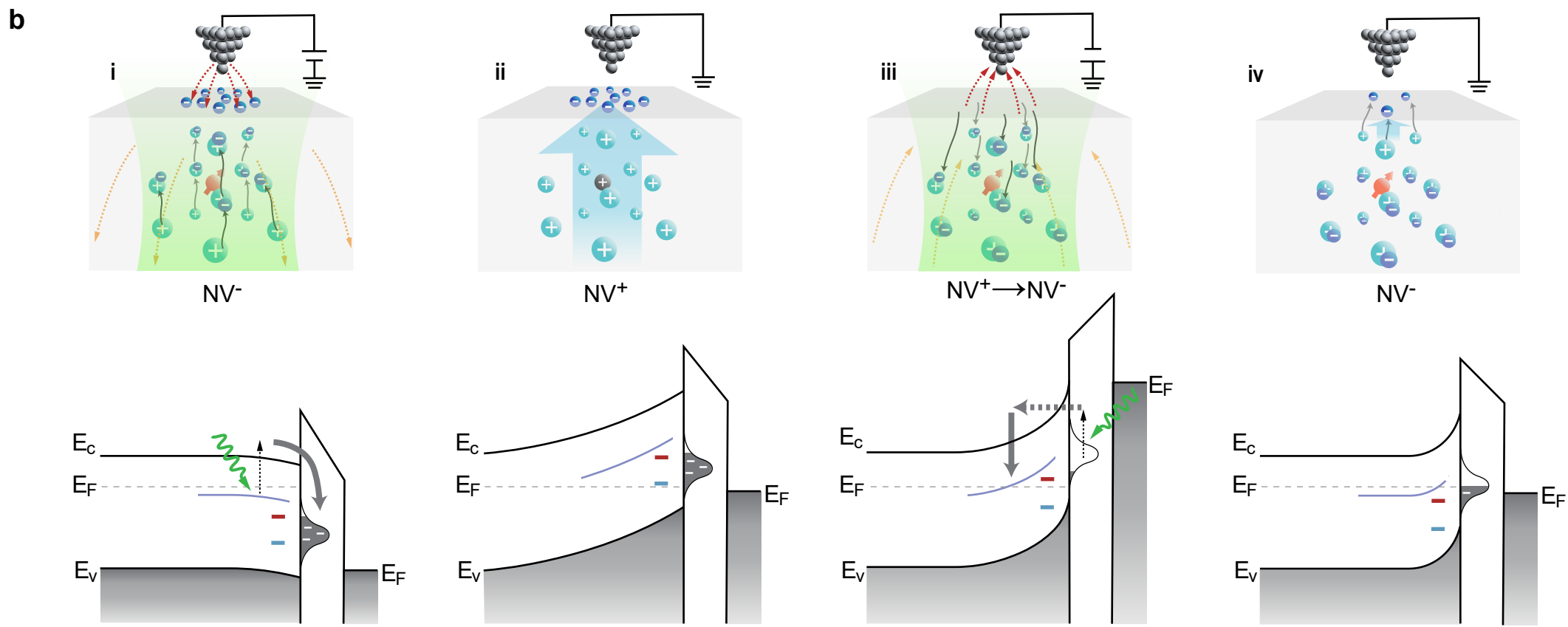

**Figure 4**

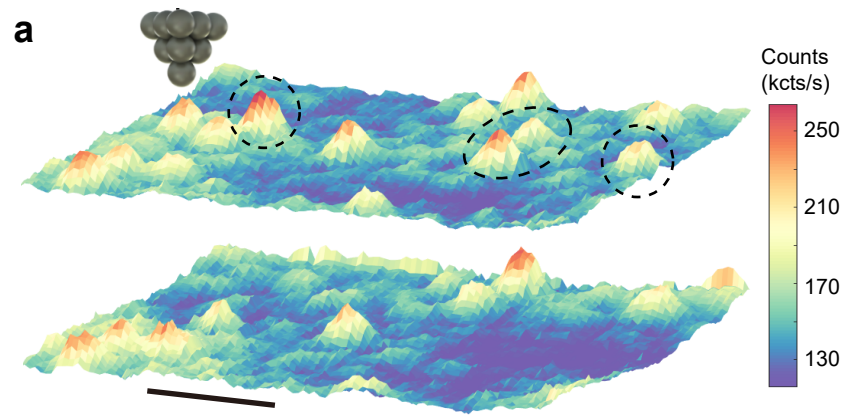
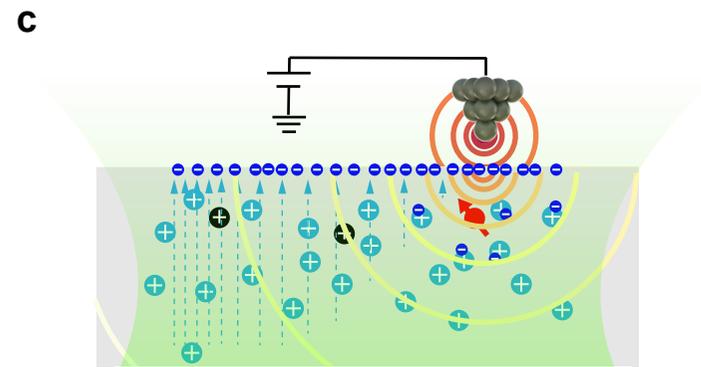
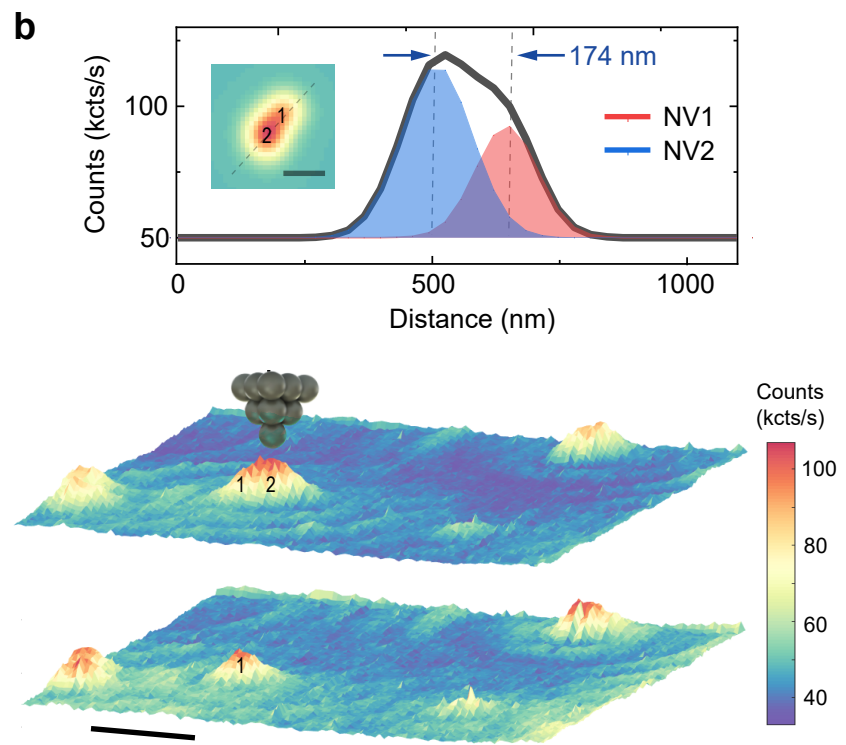
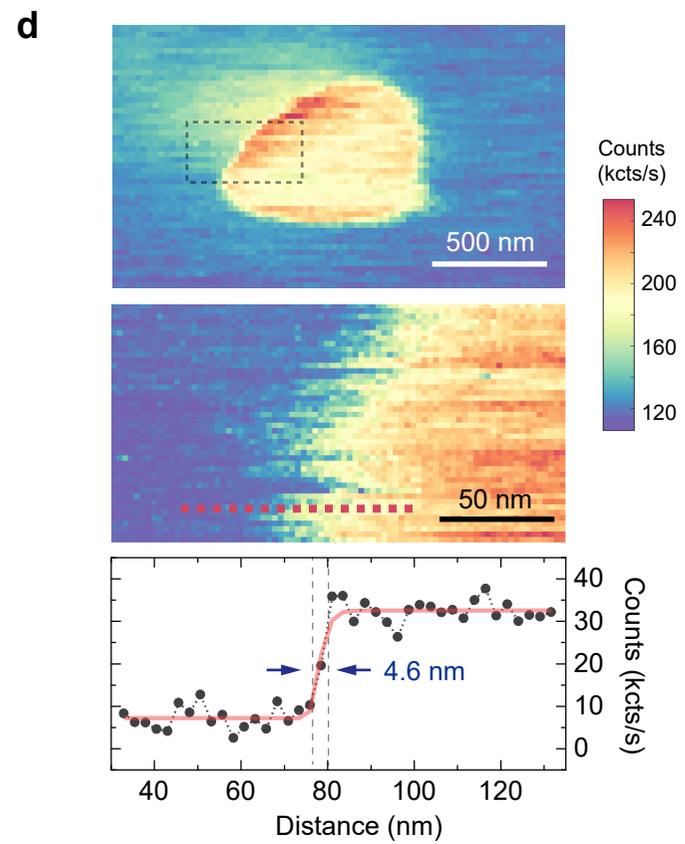

**Figure 5**